\documentclass[twocolumn,prl,showpacs,preprintnumbers,amsmath,amssymb]{revtex4}
\usepackage{graphicx,epsfig}
\usepackage{dcolumn}
\usepackage{bm}
\begin{document}
\title{Decoupling phenomena in supercooled liquids: Signatures in the energy 
landscape}
\author{Dwaipayan Chakrabarti and Biman Bagchi} 
\email{bbagchi@sscu.iisc.ernet.in}
\affiliation{Solid State and Structural Chemistry Unit, Indian Institute of 
Science, Bangalore 560012, India}

\begin{abstract}
A significant deviation from the Debye model of rotational diffusion in the 
dynamics of orientational degrees of freedom in an equimolar mixture of ellipsoids
of revolution and spheres is found to begin {\it precisely} at a temperature at
which the average inherent structure energy of the system starts falling with drop
in temperature. We argue that this onset temperature corresponds to the emergence 
of the $\alpha$-process as a distinct mode of orientational relaxation. Equally
important, we find that the coupling between the rotational and translational 
diffusion breaks down at a still lower temperature where a sharp change occurs in 
the temperature dependence of the average inherent structure energy. 
\end{abstract}


\maketitle

The relaxation phenomena in supercooled liquids continue to stimulate intense
research interests despite persistent research activity over decades 
\cite{Angell-JAP-2000, Debenedetti-Nature-2001}. A major contribution to this
activity has come from a variety of experimental techniques, e.g., dielectric 
relaxation spectroscopy \cite{DRS}, light scattering \cite{LS}, time resolved 
optical spectroscopy \cite{TROS}, NMR spectroscopy 
\cite{NMRS,Diezemann-JNCS-1998}, and optical Kerr effect spectroscopy \cite{OKES},
that probe dynamics of orientational degrees of freedom (ODOF). These experiments 
reveal an array of dynamical features some of which might owe their origin to the 
nontrivial interplay between orientational and translational degrees of freedom 
\cite{TRC}. On the other hand, a decoupling between rotational and translational 
diffusion is observed in deeply supercooled molecular liquids in the sense that 
orientational correlation time continues to track the viscosity as given by the 
Debye-Stokes-Einstein (DSE) relationship while translational diffusion coefficient
does not in contrary to what is predicted by the Debye-Einstein (DE) relation 
\cite{break,decouple}. The $\alpha-\beta$ bifurcation 
\cite{Johari-ANYAS-1976,Kivelson-JCP-1989-1991,Rossler-PRL-1992}, which 
commonly refers to the bifurcation of two peaks in the dielectric relaxation 
spectra \cite{Johari-ANYAS-1976}, marks yet another decoupling, this time between 
two distinct mechanisms for orientational relaxation in liquids composed of 
non-spherical molecules. The bifurcation temperature $T_{B}$ was believed to be 
close to the mode-coupling critical temperature $T_{c}$ \cite{Rossler-PRL-1992}, 
but it has been recently shown in broadband dielectric relaxation measurements 
that $\alpha$- and $\beta$-relaxations merge together only well above $T_{c}$ 
\cite{Fujima-PRE-2002}.

Here we address the decoupling phenomena from the perspective of potential energy 
landscape by studying a system with {\it orientational degrees of freedom} across 
the supercooled regime. The energy landscape formalism is an approach that 
explores the features of the underlying potential energy surface of a system for 
understanding its complex dynamics \cite{Goldstein-JCP-1969,Stillinger-PRA-1983,
Stillinger-Science-1995,Wales-book,Sastry-Nature-1998}. In an 
appealing landscape study of a binary mixture of spheres 
\cite{Sastry-Nature-1998}, Sastry {\it et. al.} found that the onset of 
non-exponential relaxation in the supercooled regime corresponded to the 
temperature, below which the dynamics of the system was influenced by its energy 
landscape. As in Ref. \cite{Sastry-Nature-1998}, the focus in most molecular 
dynamics simulation studies on supercooled liquids, with a few notable exceptions 
\cite{rot}, has been on {\it atomic systems} which involve only translational 
degrees of freedom (TDOF) \cite{Kob-JPCM-1999}. 

We investigate a binary mixture of $128$ Gay-Berne ellipsoids of 
revolution and $128$ Lennard-Jones spheres along an isochore at a series
of temperatures down to the deeply supercooled state \cite{Chakrabarti-PRE-2005}. 
The choice of such a system is motivated by the success of binary mixtures 
of Lennard-Jones spheres \cite{Kob-PRE-1995}, that are widely used for computer 
simulations of supercooled liquids. In our system \cite{Chakrabarti-PRE-2005}, the
interaction between the ellipsoids of revolution is given by the Gay-Berne pair 
potential \cite{Gay-JCP-1981}, which  
explicitly incorporates anisotropy in both the attractive and the repulsive parts 
of the interaction with a single-site representation for each ellipsoid of 
revolution. The interaction potential between a sphere and an ellipsoid of 
revolution, which is chosen to be a prolate (with aspect ratio $\kappa = 2$), is 
given by following Cleaver and coworkers \cite{Cleaver-PRE-JCP}. We have 
determined the set of energy and length parameters such that {\it neither any 
phase separation occurs nor any liquid crystalline phase with orientational order 
appears even at the lowest temperature studied at a high density} 
\cite{Chakrabarti-PRE-2005,binary-parameter}. Across the supercooled regime, the
diffusion coefficients for both the spheres and the ellipsoids of revolution are 
found to follow a power law temperature dependence, {\it i.e.}, 
$D_{t} = C_{D}(T-T_{c})^{\gamma_{D}}$, with $T_{c} = 0.454$ for the former and 
$T_{c} = 0.460$ for the latter. This suggests, within the error limit, the 
critical temperature $T_{c}$ to be independent of the type of particle as 
predicted by the mode-coupling theory (MCT).      
 
In Fig. 1, we show the temperature dependence of the average energy of the 
inherent structures for our binary system. At high temperatures $(T > 1.0)$, the 
average inherent structure energy remains fairly insensitive to temperature
variation. Below $T \simeq 1.0$, this energy decreases progressively up to the 
lowest temperature studied here. We find that this crossover temperature
corresponds to the onset of non-exponential relaxation in the decay of the self 
intermediate scattering function $F_{s}(k,t)$ (data not shown). This can be taken 
as a dynamical signature of the crossover behavior in translational degrees of 
freedom. Such a crossover in the average depth of the potential energy minima 
explored by the system has been observed earlier \cite{Sastry-Nature-1998}, and 
shown to be consistent with the thermal sampling of a Gaussian distribution of 
energies for the local minima \cite{gaussian,Wales-book}. 
The latter, within harmonic approximation, predicts an inverse temperature 
dependence of the average inherent structure energy \cite{gaussian}. We, 
however, observe {\it two distinct temperature regimes}, where the inverse 
temperature dependence of the average inherent structure energy holds true, 
{\it with a sharp change} at a second crossover temperature $T \simeq 0.6$ as 
illustrated in the inset of Fig. 1. We now investigate whether the temperature 
dependence of the average inherent structure energy has any correlation with 
change in the dynamics of orientational degrees of freedom.
\begin{figure}
\epsfig{file=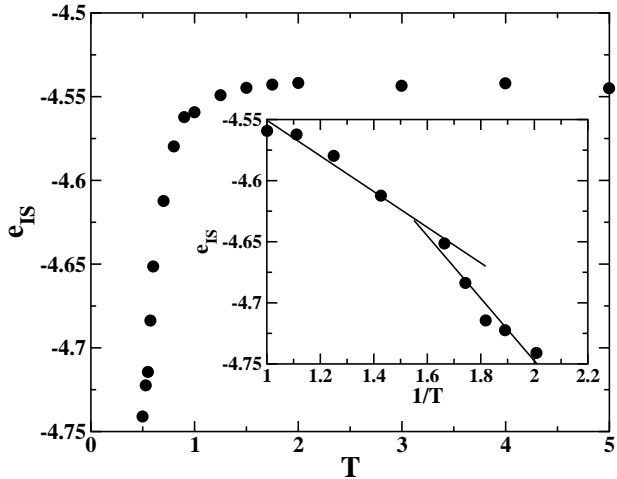,width=7cm,angle=270}
\caption{The average inherent structure energy per particle $e_{IS}$ of our binary
system as a function of temperature over the whole temperature range studied. The 
inset plots $e_{IS}$ versus $1/T$ over the temperature range across which the 
average inherent structure energy is on a decline. The solid lines are the linear 
fits to the data over two distinct temperature regimes.}   
\end{figure}

\begin{figure}
\epsfig{file=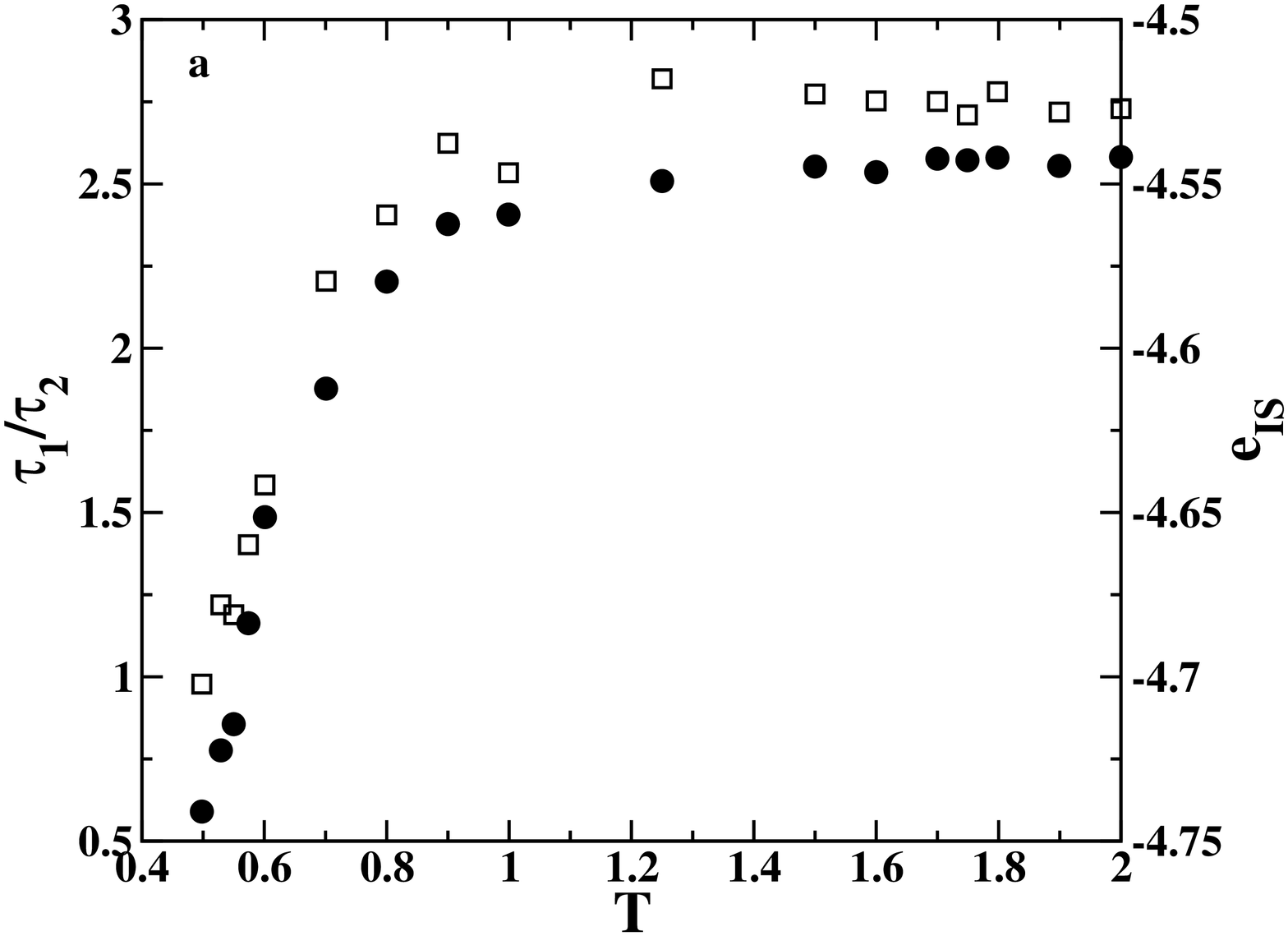,width=6cm,angle=270}
\epsfig{file=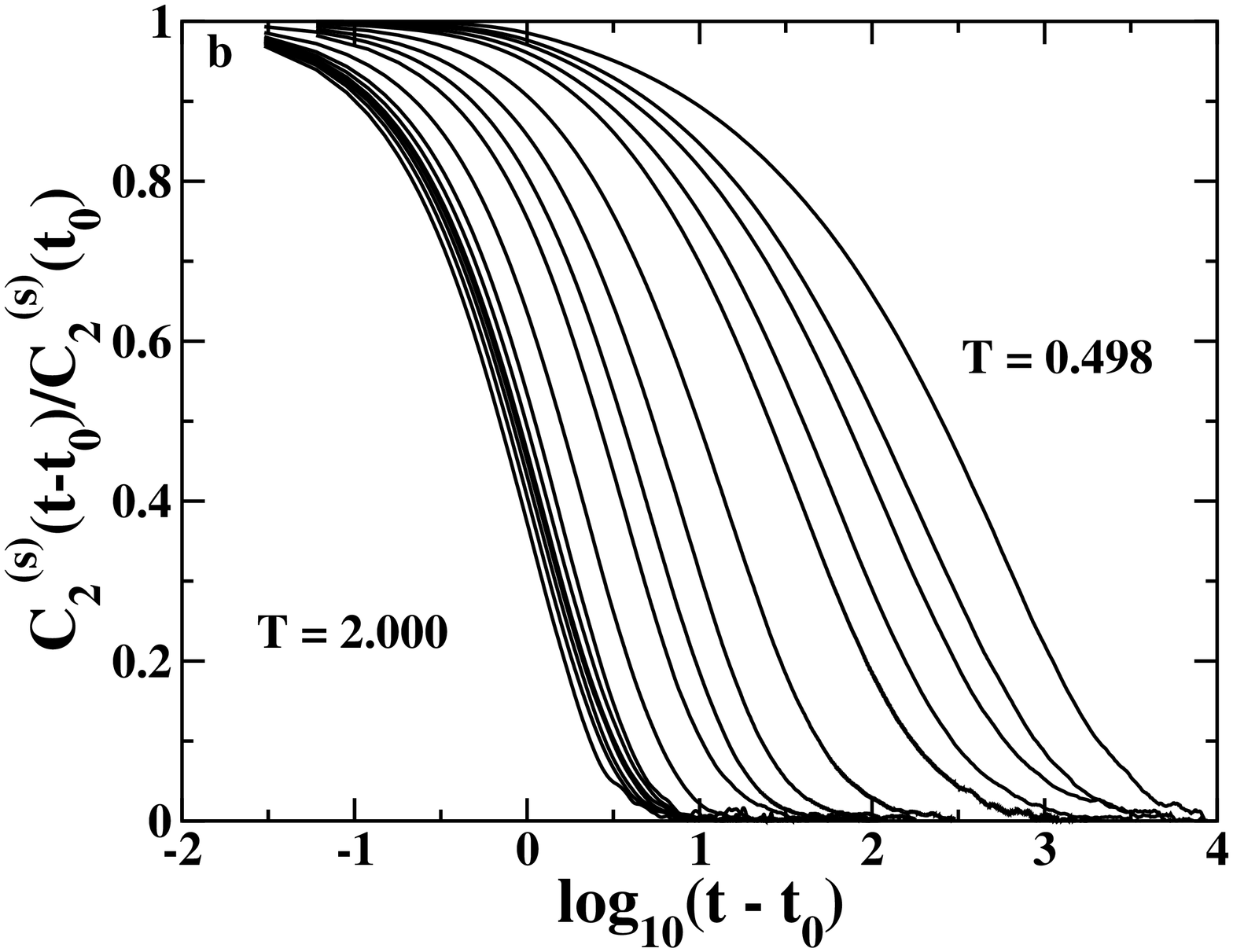,width=6cm,angle=270}
\epsfig{file=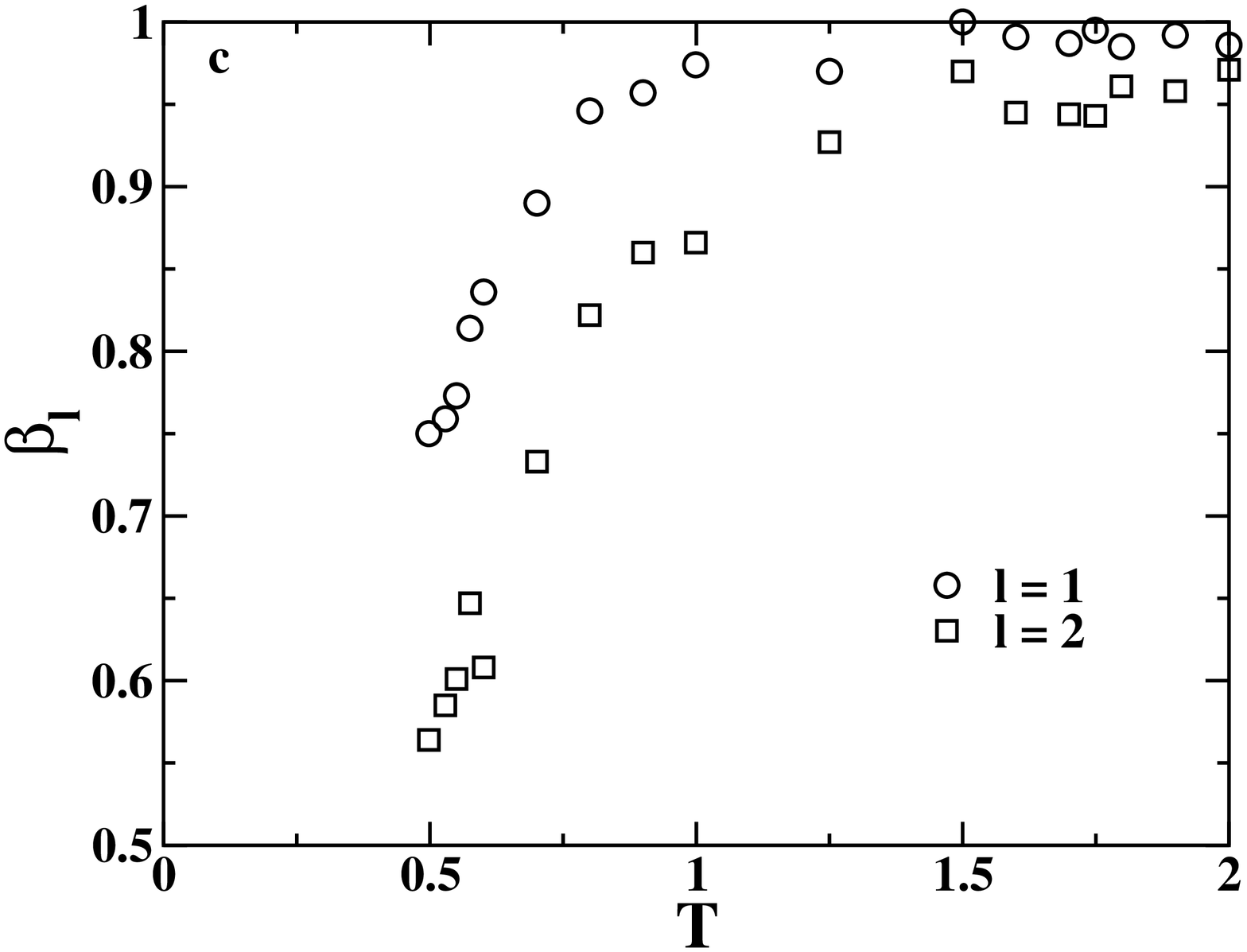,width=6cm,angle=270}
\caption{{\bf {\large a.}} The temperature dependence of 
$\tau_{1}(T)/\tau_{2}(T)$, the ratio of the first to second rank rotational 
correlation times (squares). On a different scale (appearing on the right) shown 
again is the temperature dependence of the average inherent structure energy per 
particle $e_{IS}$ for the purpose of comparison (circles). {\bf {\large b.}} The 
time evolution of the single-particle second-rank orientational correlation 
function at several temperatures with a shift in the time origin to 
$t_{0} = 1.2$ followed by normalization to the corresponding value at $t_{0}$. 
{\bf {\large c.}} The exponents values $\beta_{1}(T)$ and 
$\beta_{2}(T)$, obtained from the fits to the stretched exponential form of the 
first and second rank single-particle orientational correlation functions, 
respectively, as a function of temperature.}
\end{figure}
Figure 2a shows how the ratio of the first to second rank rotational correlation 
times, $\tau_{1} / \tau_{2}$, evolves as temperature drops. The $l$-th rank 
rotational correlation time $\tau_{l}$ is defined as 
$\tau_{l} = \int_{0}^{\infty} C_{l}^{(s)}(t) dt$, 
where $C_{l}^{(s)}(t)$ is the $l$-th rank single-particle orientational 
correlation function:
\begin{equation}
C_{l}^{(s)}(t)={\frac{\langle{\sum_{i} P_{l}({\bf \hat e}_{i}(0)\cdot
{\bf \hat e}_{i}(t))}\rangle}{
\langle{\sum_i P_{l}({\bf \hat e}_{i}(0)\cdot {\bf \hat e}_{i}(0))}\rangle}},
\label{cs2}
\end{equation}
Here ${\bf \hat e}_{i}$ is the unit vector along the principal axis of
symmetry of the ellipsoid of revolution $i$, $P_{l}$ is the $l${\it th}
rank Legendre polynomial and the angular brackets stand for ensemble averaging. It
is evident in Fig. 2a that the ratio has a value close to $3$ as
predicted by the Debye model of rotational diffusion at high temperatures 
$(T > 1.0)$ and {\it starts declining steadily from $T \simeq 1.0$ 
until it reaches a value nearly unity at low temperatures}. In the diffusive 
limit, the orientational motion occurs in small steps, while a value for this 
ratio close to $1$ is taken to suggest that the orientational motion takes place 
via long angular jumps \cite{Kivelson-JCP-1989-1991}.

\begin{figure}
\epsfig{file=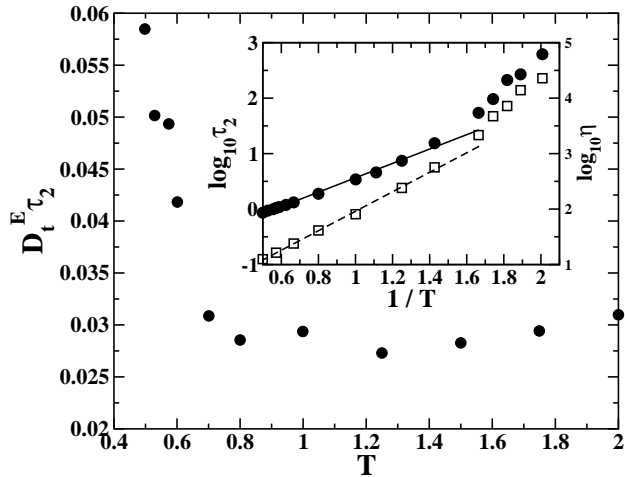,width=7cm,angle=270}
\caption{The product of the translational diffusion coefficient 
$D_{t}^{E}$ and the second-rank rotational correlation time $\tau_{2}$ for the 
ellipsoids of revolution as a function of temperature. Inset: The inverse 
temperature dependence the logarithm of the second-rank rotational correlation 
time $\tau_{2}$ (circles). On a different scale (appearing on the right of the
inset) shown is the inverse temperature dependence the logarithm of the shear 
viscosity (squares). The solid and dashed lines are the respective Arrhenius fits 
to data over a restricted temperature range.}
\end{figure}
\begin{figure}
\epsfig{file=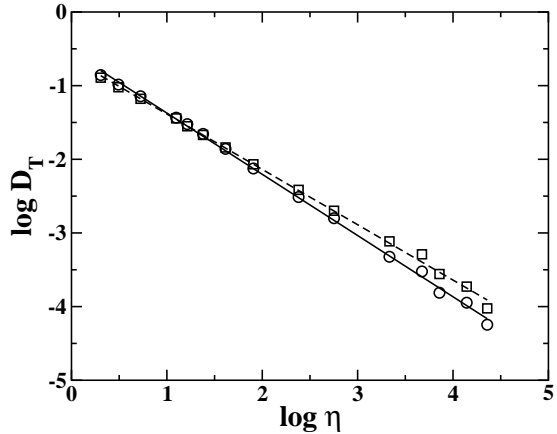,width=7cm,angle=270}
\caption{The translational diffusion coefficient $D_{t}$ versus the shear
viscosity $\eta$ in a log-log plot for both spheres (circles) and the ellipsoids 
of revolution (squares). The solid and dashed lines are the respective linear 
fits.}
\end{figure}
At high temperatures, the long time decay of $C_{2}^{(s)}(t)$ is exponential while
the Kohlrausch-Williams-Watts stretched exponential form provides a reasonable fit
to the long-time behavior at low temperatures. To eliminate
the short-time Gaussian time dependence, we show in Fig. 2b the time evolution of 
the function $C_{2}^{(s)}(t-t_{0})/C_{2}^{(s)}(t_{0})$ for $t > t_{0}$, and 
consider the stretched exponential form $exp[-((t-t_{0})/\tau(T))^{\beta_{2}(T)}]$
that takes into account this transformation to fit the data. The deviation of the 
exponent $\beta_{2}(T)$ $(0 < \beta_{2}(T) < 1)$ from unity is a measure of the 
degree of non-exponential relaxation. Figure 2c shows the temperature dependence 
$\beta_{2}(T)$ and also $\beta_{1}(T)$, the latter corresponding to the long-time 
decay of $C_{1}^{(s)}(t)$ (not shown here). While both $\beta_{1}(T)$ and 
$\beta_{2}(T)$ are very close to unity at high temperatures, they start dropping 
as temperature falls. It is evident that {\it the stretching is more pronounced 
in the second rank orientational correlation function as compared to the first 
rank orientational correlation function}. Such an observation has been reported 
previously while comparing data from dielectric and NMR experiments 
\cite{Diezemann-JNCS-1998}. We note that the signature of non-exponential 
relaxation in $C_{2}^{(s)}(t)$ {\it first becomes appreciable} and later gets 
progressively more pronounced as temperature drops below the onset temperature 
$T \simeq 1.0$. 

It follows from above that the onset of the growth of the depth of the potential 
energy minima explored by the system correlates with a {\it change in the 
mechanism of orientational motion from being simply diffusive}. There is 
evidence for the $\beta$-relaxation to be diffusive in character and 
the orientational relaxation above $T_{B}$ is associated with the $\beta$-process
only \cite{Kivelson-JCP-1989-1991}. Thus, the onset temperature corresponds to the
emergence of the $\alpha$-process as a distinct mode of orientational relaxation 
and {\it appears to be coinciding with the bifurcation temperature} $T_{B}$. We, 
however, find that the latter is 
somewhat higher than $T_{c}$ as indeed observed in Ref. \cite{Fujima-PRE-2002} in 
contrary to what often believed. Nevertheless, Fig.2a illustrates that the decline
of the ratio $\tau_{1} / \tau_{2}$ {\it closely tracks the fall of 
the average inherent structure energy of the system}. Such a correlation, to the 
best of our knowledge, has not been demonstrated before. Stillinger
interpreted the $\alpha-\beta$ bifurcation in terms of the topography of the 
potential energy landscape \cite{Stillinger-Science-1995}. In Stillinger's 
picture, the $\beta$-processes correspond to the elementary relaxations between 
contiguous basins while the $\alpha$-processes invoke escape from one metabasin 
and eventually into other with an involvement of high free energy of activation. 
Such a description is consistent with the growth of the depth of the potential 
energy minima explored by the system below the bifurcation temperature.     

We now address the decoupling between rotational and translational diffusion.
The combination of the SE and DSE equations predicts the product $D_{t}\tau_{2}$
to be independent of temperature even when the
macroscopic observable viscosity increases by many orders of magnitude as the 
glass transition temperature $T_{g}$ is approached from above \cite{break}. 
Figure 3 shows that such a relationship breaks down at $T \simeq 0.6$ and below 
with the product growing fast with decrease in temperature. The inset of Fig. 3 
illustrates that the decoupling between the two microscopic observables occurs 
precisely at the same temperature at which both the orientational correlation time
$\tau_{2}$ and the viscosity start showing steady deviation from the Arrhenius 
temperature behavior. The inset of Fig. 1 shows that at this temperature the 
linear variation of the average inherent structure energy with the inverse 
temperature undergoes a {\it sharp change with an increase in the rate of fall}.
   
In Fig. 4, we show the variation of the translational diffusion coefficient 
$D_{t}$ with the coefficient of shear viscosity $\eta$ in a log-log plot over the 
whole temperature range studied here. The linearity of the curve implies a power 
law dependence: $D_{t} \propto \eta^{-\alpha}$, $\alpha$ being the exponent, for 
both the spheres and the ellipsoids of revolution. We find $\alpha = 0.83$ for the
former and $\alpha = 0.75$ for the latter. The {\it fractional power law 
dependence} suggests the enhancement of translational diffusion relative to what 
the SE relationship predicts. The $\alpha$ values obtained here compare well with
$0.77$, observed by Ediger and coworkers in a direct measurement of self-diffusion
of a single-component glass-forming liquid reported recently 
\cite{Swallen-PRL-2003}.  

We now argue that the deviation from the Debye behavior of rotational diffusion
is due to a crossover from a collision dominated to a correlation dominated regime
of orientational relaxation. Below $T\simeq 1$, the equilibrium pair correlations 
are expected to grow as $e_{IS}$ starts its descend. The effects of correlations 
can be included via the standard Zwanzig-Mori continued fraction representation:
$\hat C_{l}(z) = 1/(z+l(l+1)k_{B}T/(I(z+ \hat \Gamma_{l}(z)))$,
where the memory kernel $\hat \Gamma_{l}(z)$ can be decomposed in the spirit of 
MCT as $\hat \Gamma_{l}(z) \simeq \hat \Gamma_{coll}(z) + 
\Delta \hat \Gamma_{l}(z)$, the effect of correlations being contained in 
$\Delta \hat \Gamma_{l}(z)$, and $z$ stands for the Laplace frequency. The rank 
dependence of the memory function can be approximately expressed in terms of the 
torques-torque time correlation function \cite{rank}. Because of the up-down 
symmetry of the ellipsoids of revolution, the contribution of the correlation to 
$\Delta \hat \Gamma_{2}$ is expected to be larger than that to 
$\Delta \hat \Gamma_{1}$, particularly at $t = 0$. Thus, $\hat \Gamma_{2} > 
\hat \Gamma_{1}$ and this would lower the value of the ratio $\tau_{1}/\tau_{2}$. 
An opposite trend has been observed for dipolar systems where 
$\hat \Gamma_{1} > \hat \Gamma_{2}$ and an upward deviation for the ratio 
$\tau_{1}/\tau_{2}$ is observed \cite{rank}.

In summary, we have established a correlation of the breakdown of the Debye
model of orientational relaxation and the $\alpha-\beta$ bifurcation with the
manner of exploration of the underlying potential energy landscape in a model 
system. Equally important, the decoupling between the rotational and 
translational diffusion is signaled by a sharp rise in the rate of fall of 
the average inherent structure energy with the inverse temperature.
   
We thank M. Ediger and C. A. Angell for correspondence. This work was supported in
parts by grants from DST and CSIR, India. DC acknowledges UGC, India for providing
Research Fellowship.


\begin{references}
\bibitem{binary-parameter} All quantities are given in reduced units, 
defined in terms of the Lennard-Jones potential parameters $\epsilon_{LJ}$ and
$\sigma_{LJ}$: length in units of $\sigma_{LJ}$, temperature in units of 
$\epsilon_{LJ}/k_{B}$, $k_{B}$ being the Boltzmann constant, and time in units of
$(\sigma_{LJ}^{2}m/\epsilon_{LJ})^{1/2}$, $m$ being the mass of spheres. We set
the molecular masses of both spheres and ellipsoids and the moment of inertia of 
the ellipsoids of revolution equal to unity. The various energy and length 
parameters of the interaction potentials for our binary mixture were chosen as 
follows: $\epsilon_{LJ} = 1.0$ , $\epsilon_{GB} = 0.5$, $\epsilon_{SE} = 1.5$, 
$\sigma_{LJ} = 1.0$, $\sigma_{GB} = 1.0$, and $\sigma_{SE} = 1.0$. See Ref. 
\cite{Cleaver-PRE-JCP} for the general implication of the symbols. Unlike in
Ref. \cite{Cleaver-PRE-JCP}, here the subscripts are used to distinguish 
between the Lennard-Jones (LJ) interactions, the Gay-Berne (GB) 
interactions, and the sphere-ellipsoid (SE) interactions in the binary 
mixture. For the Gay-Berne interactions, $\kappa = 2, \kappa^{\prime} = 5, 
\mu = 2$, and $\nu = 1$. For the sphere-ellipsoid interaction 
$\epsilon_{E} / \epsilon_{S}  = 0.2$ and $\mu = 2$. With the set of 
parameter values, the binary mixture was investigated along an isochore
corresponding to the reduced density $\rho = 0.8$ at a series of reduced 
temperatures ranging from $4.997$ to $0.498$. The production runs were in 
microcanonical ensembles and the results presented here are average over three
independent runs.
\end{references}
\end{document}